\documentclass[twocolumn,showpacs,preprintnumbers,amsmath,amssymb,superscriptaddress]{revtex4}

\usepackage{graphicx}
\usepackage{bm}
\usepackage{scalefnt}

\usepackage{color}

\newcommand{\Lmu}{L_\mu}

\allowdisplaybreaks

\begin{document}


\preprint{DESY-13-253, LPN14-003, 
  SFB/CPP-14-03, SI-HEP-2014-01, TTP14-002, TTK-14-01, TUM-HEP-922/14}
\title{Three-loop matching of the vector current}
\author{Peter Marquard}
\affiliation{Institut f\"ur Theoretische Teilchenphysik, Karlsruhe
  Institute of Technology (KIT), D-76128 Karlsruhe, Germany}
\affiliation{Deutsches Elektronen Synchrotron DESY,
      Platanenallee 6, D-15738 Zeuthen, Germany}

\author{Jan H. Piclum}
\affiliation{Institut f\"ur Theoretische Teilchenphysik und Kosmologie, RWTH
  Aachen, D-52056 Aachen, Germany}
\affiliation{Physik-Department T31, Technische Universit\"at M\"unchen,
  D-85748 Garching, Germany}

\author{Dirk Seidel}
\affiliation{Theoretische Physik 1, Universit\"at Siegen, D-57068 Siegen, Germany}

\author{Matthias Steinhauser}
\affiliation{Institut f\"ur Theoretische Teilchenphysik, Karlsruhe
  Institute of Technology (KIT), D-76128 Karlsruhe, Germany}

\date{\today}

\begin{abstract}
We evaluate the three-loop corrections to the matching coefficient 
of the vector current between Quantum Chromodynamics (QCD) 
and non-relativistic QCD.
The result is presented in the $\overline{\rm MS}$ scheme where large 
perturbative corrections are observed.
The implications on the threshold production of top quark pairs
are briefly discussed.
\end{abstract}


\maketitle



\section{Introduction}

In the recent years effective field theories constructed from Quantum
Chromodynamics (QCD) have been enormously successful to
describe phenomena where masses and momenta follow certain limits.
Among them is non-relativistic QCD (NRQCD)~\cite{Caswell:1985ui,Bodwin:1994jh}
which is applicable to a system of two heavy quarks moving with small relative
velocity. Next to properties of the
$\psi$ and $\Upsilon$ families also the threshold production of top quark
pairs is among the prominent examples (see, e.g., Ref.~\cite{Brambilla:2004wf}
for a review). 

The common method to construct an effective theory is based on the
so-called matching procedure: appropriately chosen Green's functions are
computed in the full and effective theory and equality is required up to
power-suppressed terms. In this way the couplings of the effective operators
(i.e. the matching coefficients) are determined which completely specifies the
effective theory.

A crucial operator both in QCD and NRQCD is the vector current of a heavy
quark-antiquark pair. The corresponding matching coefficient enters as a
building block in a variety of physical observables, for example the
bottom-quark mass from $\Upsilon$ sum rules (see, e.g.,
Refs.~\cite{Pineda:2006gx,Hoang:2012us} 
for recent analyses) and top-quark threshold
production at a future electron positron linear
collider~\cite{Hoang:2000yr}. The latter process allows for an extraction of
the top-quark mass with an accuracy below
100~MeV~\cite{Martinez:2002st,Seidel:2013sqa,Horiguchi:2013wra} --- an
improvement of about an order of magnitude as compared to the current
results from the Fermilab Tevatron or the the CERN Large Hadron
Collider~\cite{Beringer:1900zz}.

Several quantities are needed in order to perform a third-order analysis
of a heavy quark-antiquark system at threshold. Ultrasoft effects have been
considered in Refs.~\cite{Beneke:2007pj,Beneke:2008cr}, the three-loop static
potential has been 
computed in Refs.~\cite{Smirnov:2008pn,Smirnov:2009fh,Anzai:2009tm} and in
Ref.~\cite{Beneke:2007uf,Beneke:2008ec} 
a preliminary analysis of the top-quark threshold
production cross section has been presented including also third-order
potential effects. Details on the potential contributions can
be found in Refs.~\cite{BenKiy_I,BenKiy_II}.
In this paper we compute the three-loop matching coefficient between
the vector current in
QCD and NRQCD. Thus all ingredients are available to obtain
the complete next-to-next-to-next-to leading 
order QCD prediction of the cross section $e^+e^-\to t\bar{t}$ close to
threshold or the decay width of the $\Upsilon(1S)$ meson to leptons. 
The results for the latter are presented in an accompanying
paper~\cite{bottom_NNNLO}
where all building blocks are combined to a phenomenological analysis.


\section{Vector currents in QCD and NRQCD}

The vector current in the full theory (QCD) is given by
\begin{eqnarray}
  j_v^\mu &=& \bar{Q} \gamma^\mu Q\,,
\end{eqnarray}
where $Q$ denotes a generic heavy quark with mass $m_Q$.
On the other hand in the effective theory (NRQCD) the current is represented by
an expansion in $1/m_Q$ where at each order effective operators have to be
considered which are multiplied by
coefficient functions. The leading contribution involves one operator given by
\begin{eqnarray}
  \tilde{j}^k = \phi^\dagger \sigma^k \chi\,,
\end{eqnarray}
where
$\phi$ and $\chi$ are two-component Pauli spinors for quark and
anti-quark, respectively, and $\sigma^k$ ($k=1,2,3$) are the Pauli matrices.
Hence, the matching coefficient of the vector current 
is defined through
\begin{eqnarray}
  j^k_v &=& c_v(\mu) \tilde{j}^k 
  + {\cal O}\left(\frac{1}{m_Q^2}\right)
  \label{eq::def_of_cv}
  \,.
\end{eqnarray}
Note that the 0-component of $j_v^\mu$ is only relevant for the
power-suppressed contributions.

The purpose of this paper is the evaluation of the purely gluonic
three-loop corrections to $c_v$. The fermionic contributions have already been
considered in Refs.~\cite{Marquard:2006qi,Marquard:2009bj}. 

In order to compute $c_v$ it is convenient to consider on-shell vertex
corrections involving the currents $j^k_v$ and $\tilde{j}^k$. After taking
into account the wave function renormalization one obtains 
(see also Ref.~\cite{BenKiy_I})
\begin{eqnarray}
  Z_2 \Gamma_v &=& c_v \tilde{Z}_2 \tilde{Z}_v^{-1} \tilde{\Gamma}_v
  + \ldots
  \,,
  \label{eq::def_cv}
\end{eqnarray}
where the quantities with a tilde are defined in the effective theory and 
the ellipsis represents terms suppressed by the heavy-quark mass.
$\tilde{Z}_v^{-1}$ is the renormalization constant of the current
$\tilde{j}^k$ which is used to subtract the remaining poles after
renormalization. These poles are due to the separation of long and short
distance contributions in the effective theory. In order to evaluate
physical quantities it is important that the same subtraction scheme is
also adopted in the contributions originating from the effective
theory~\cite{Beneke:2007pj,Beneke:2008cr}. It is well-known that in the
full theory the renormalization constant of the vector current is
equal to one.

\begin{figure}[t]
  \includegraphics[width=0.7\columnwidth]{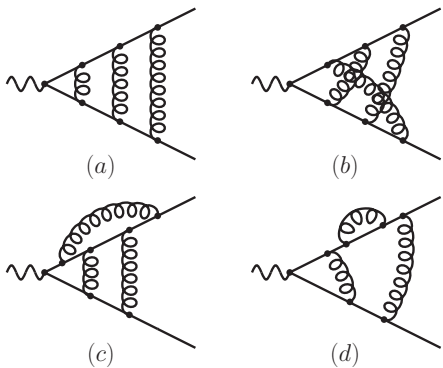}
\caption{\label{fig::sample} Feynman diagrams contributing to $\Gamma_v$.
  Straight and curly lines denote heavy quarks with mass $m_Q$ and gluons,
  respectively.}
\end{figure}

In Eq.~(\ref{eq::def_cv}) $Z_2$ is the on-shell wave function renormalization
constant which has been computed up to three-loop accuracy in
Refs.~\cite{Broadhurst:1991fy,Melnikov:2000zc,Marquard:2007uj}. 
$\Gamma_v$ denotes the one-particle irreducible vertex diagrams
with on-shell quarks carrying momenta $q_1$ and $q_2$. It incorporates all
one-particle irreducible vertex graphs and the corresponding counterterms 
for $m_Q$ and $\alpha_s$. Sample Feynman diagrams are shown in
Fig.~\ref{fig::sample}.  

The counterparts of $\Gamma_v$ and $Z_2$ in the effective theory can be found
on the right-hand side of Eq.~(\ref{eq::def_cv}). It is convenient to apply
the threshold expansion~\cite{Beneke:1997zp,Smirnov:2013} to
Eq.~(\ref{eq::def_cv}). This requires the identification of the hard, soft,
potential and ultrasoft momentum regions in the integrals contributing to
$\Gamma_v$ and $\tilde{\Gamma}_v$. Since NRQCD is obtained from QCD by
integrating out the hard modes one has by construction that the soft,
potential and ultrasoft modes agree in $\Gamma_v$ and $\tilde{\Gamma}_v$ and
thus drop out from Eq.~(\ref{eq::def_cv}).  As a consequence $\Gamma_v$ is
evaluated for $q^2=(q_1+q_2)^2=4m_Q^2$, which corresponds to the leading term
of the hard integration region, and $\tilde{\Gamma}_v=1$. Furthermore, we have
$\tilde{Z}_2=1$.

There are several technical difficulties which one has to overcome in order to
compute the vertex corrections. Among them are the large number of diagrams
which leads to several thousand Feynman integrals to be evaluated in the first
place, their reduction to a small set of about 100 basis integrals, so-called
master integrals, and the evaluation of the latter in an expansion in
$\epsilon=(4-D)/2$, where $D$ is the space-time dimension. The last two tasks
become more complicated by the additional condition $q^2=4m_Q^2$ on the
external momentum.  An automated setup for the calculation has been described
in Ref.~\cite{Marquard:2009bj} and applied to the fermionic contributions.
Its core parts are a powerful implementation of Laporta's algorithm in the
program {\tt CRUSHER}~\cite{PMDS}, and {\tt
  FIESTA}~\cite{Smirnov:2008py,Smirnov:2009pb,Smirnov:2013eza} which is based
on sector 
decomposition and is used for the numerical integration of the master
integrals.  The main differences to the gluonic part considered in this paper
are the larger number of diagrams and the increased complexity of the integrals
which have to be reduced to master integrals. Furthermore, the master integrals
are more numerous and more involved.


\section{\label{sec::match}Matching coefficient to order \boldmath$\alpha_s^3$}

Before discussing the matching coefficient it is instructive to 
consider the renormalization constant $\tilde{Z}_v$. The analytical results can
be extracted from
Refs.~\cite{Beneke:1997jm,Marquard:2006qi,Kniehl:2002yv,Beneke:2007pj}. 
\begin{eqnarray}
  \lefteqn{\tilde{Z}_v =}
  \nonumber\\&&
  1 + \left(\frac{\alpha_s^{(n_l)}(\mu)}{\pi}\right)^2 
  \frac{C_F\pi^2}{\epsilon}\left(
    \frac{1}{12} C_F + \frac{1}{8} C_A \right) 
  \nonumber \\ &&\mbox{}
  + \left(\frac{\alpha_s^{(n_l)}(\mu)}{\pi}\right)^3  C_F\pi^2
  \nonumber\\ &&\mbox{}
  \times \Bigg\{ C_F^2 \left[\frac{5}{144\epsilon^2} + \left(
    \frac{43}{144} - \frac{1}{2} \ln2 + \frac{5}{48} \Lmu
    \right)\frac{1}{\epsilon} \right] \nonumber \\ &&\mbox{}
  +C_FC_A \left[\frac{1}{864\epsilon^2} + \left( \frac{113}{324} +
    \frac{1}{4} \ln2 + \frac{5}{32} \Lmu \right) \frac{1}{\epsilon}
    \right] \nonumber \\ &&\mbox{}
  +C_A^2 \left[-\frac{1}{16\epsilon^2} + \left( \frac{2}{27} +
    \frac{1}{4} \ln2 + \frac{1}{24} \Lmu \right) \frac{1}{\epsilon}
    \right] \nonumber \\ &&\mbox{}
  + T n_l \left[ C_F\left(
      \frac{1}{54\epsilon^2}  
      -\frac{25}{324\epsilon}
    \right)
    + C_A \left(
      \frac{1}{36\epsilon^2}  
      - \frac{37}{432\epsilon} \right)
     \right] \nonumber \\ &&\mbox{}
  + C_F T n_h \frac{1}{60\epsilon} 
  \Bigg\} + {\cal O}(\alpha_s^4)\,,
  \label{eq::Zv}
\end{eqnarray}
where $C_A=N_c$, $C_F=(N_c^2-1)/(2N_c)$ and $T=1/2$ for a ${\rm SU}(N_c)$ gauge
group and $\Lmu=\ln(\mu^2/m_Q^2)$.
Note that the strong coupling is defined in the effective theory with $n_l$
active quarks where $n_l+n_h$ is the total number of quark flavors.
In our case we have $n_h=1$, however, we keep $n_h$ in the formulae for
convenience.

From our calculation we can extract the renormalization constant $\tilde{Z}_v$
and compare with Eq.~(\ref{eq::Zv}).  The central values of our numerical
coefficients agree at the per cent level with the analytical result of
Eq.~(\ref{eq::Zv}) which constitutes a first non-trivial check and provides
quite some confidence in the overall set-up of our calculation.

We write the perturbative expansion of the matching coefficient in the form
\begin{eqnarray}
  c_v &=& 1 + \frac{\alpha_s^{(n_l)}(\mu)}{\pi} c_v^{(1)}
  + \left(\frac{\alpha_s^{(n_l)}(\mu)}{\pi}\right)^2 c_v^{(2)}
  \nonumber\\&&\mbox{}
  + \left(\frac{\alpha_s^{(n_l)}(\mu)}{\pi}\right)^3 c_v^{(3)}
  + {\cal O}(\alpha_s^4)
  \,,
  \label{eq::cvdef}
\end{eqnarray}
and decompose $c_v^{(3)}$ according to the color structures as
\begin{eqnarray}
  \lefteqn{c_v^{(3)} =}
  \nonumber\\&&\mbox{}
  C_F \big[ C_F^2 c_{FFF} + C_F C_A c_{FFA} + C_A^2 c_{FAA}
  \nonumber\\&&\mbox{}
  + T n_l\left(
  C_F c_{FFL} + C_A c_{FAL} + T n_h c_{FHL} + T n_l c_{FLL}
  \right)
  \nonumber\\&&\mbox{}
  + T n_h\left(
  C_F c_{FFH} + C_A c_{FAH} + T n_h c_{FHH}
  \right) \big]
  \nonumber\\&&\mbox{}
  + \mbox{singlet terms}
  \label{eq::cv3ldef}
  \,.
\end{eqnarray}
Note that all color factors of the non-singlet part
can be expressed in terms of $C_F$, $C_A$, and $T$.
In this paper we do not consider the singlet contribution where the external
current does not couple to the fermion line in
the final state. At two-loop order such contributions have been
computed~\footnote{Due to Furry's theorem the two-loop singlet contribution is
  zero for the vector current.}
for axial-vector, scalar and pseudo-scalar currents in
Ref.~\cite{Kniehl:2006qw}. Their numerical effect in those cases is below 3\%
as compared to the non-singlet contributions and thus quite small.

Whereas the one-~\cite{KalSar} and
two-loop~\cite{Czarnecki:1997vz,Beneke:1997jm,Kniehl:2006qw} terms have
been known since long
the fermionic three-loop corrections became 
available only a few years ago~\cite{Marquard:2006qi,Marquard:2009bj}. 
The so-called renormalon contribution, consisting of the
one-loop diagram with arbitrary many massless quark loop insertions
in the gluon propagator, has been computed in Ref.~\cite{Braaten:1998au}.
Supersymmetric one-loop corrections to $c_v$ have been computed in
Ref.~\cite{Kiyo:2009ih}.

In the following we present the results for the individual coefficients in
Eq.~(\ref{eq::cv3ldef}) parameterized in terms of $\alpha_s^{(n_l)}(m_Q)$. The
reconstruction of the full dependence on the renormalization scale is
straightforward; the corresponding expressions can be obtained
from~\cite{webpage}.
Our results read~\footnote{Note 
  that in Ref.~\cite{Marquard:2006qi} the result has been expressed in terms
  of the coupling defined in the full theory whereas here we use the effective
  one denoted by $\alpha_s^{(n_l)}$.
  This explains the difference in the logarithmic part of the coefficient $c_{FHL}$.}
\begin{eqnarray}
  c_v^{(1)}&=&-2C_F\,,
  \nonumber\\
  c_v^{(2)}&=&\left(-\frac{151}{72}
    +\frac{89}{144}\pi^2
    -\frac{5}{6}\pi^2\ln2-\frac{13}{4}\zeta(3)\right)C_AC_F
  \nonumber\\&&\mbox{}
  +\left(\frac{23}{8}-\frac{79}{36}\pi^2
    +\pi^2\ln2-\frac{1}{2}\zeta(3)\right)C_F^2
  \nonumber\\&&\mbox{}
  +\left(\frac{22}{9}-\frac{2}{9}\pi^2\right)C_FTn_h
  +\frac{11}{18}C_FTn_l
  \nonumber\\&&\mbox{}
  -\frac{1}{2}\pi^2\left(\frac{1}{2}C_A 
    + \frac{1}{3}C_F\right)C_F\Lmu
  \,,
  \nonumber\\
  c_{FFF} &=& 36.55(0.53)
  \nonumber\\&&\mbox{}
  + \left( -\frac{9}{16} + \frac{3}{2} \ln2 \right) \pi^2 \Lmu -
  \frac{5}{32} \pi^2 \Lmu^2
  \,,  \nonumber\\
  c_{FFA} &=& -188.10(0.83)
  \nonumber\\&&\mbox{}
  + \left( -\frac{59}{108} - \frac{3}{4} \ln2 \right) \pi^2 \Lmu -
  \frac{47}{576} \pi^2 \Lmu^2
  \,,  \nonumber\\
  c_{FAA} &=& -97.81(0.38)
  \nonumber\\&&\mbox{}
  + \left( -\frac{2}{9} - \frac{3}{4} \ln2 \right) \pi^2 \Lmu +
  \frac{1}{6} \pi^2 \Lmu^2
  \,,  \nonumber\\
  c_{FFL} &=& 46.691(0.006) + \frac{25}{108} \pi^2 \Lmu -
  \frac{1}{18} \pi^2 \Lmu^2
  \,, \nonumber\\
  c_{FAL} &=& 39.624(0.005) + \frac{37}{144} \pi^2 \Lmu -
  \frac{1}{12} \pi^2 \Lmu^2
  \,, \nonumber\\
  c_{FHL} &=& -\frac{557}{162} + \frac{26}{81} \pi^2\,,
  \nonumber\\
  c_{FLL} &=& -\frac{163}{162} - \frac{4}{27} \pi^2\,,
  \nonumber\\
  c_{FFH} &=& -0.846(0.006) - \frac{1}{20} \pi^2 \Lmu\,,
  \nonumber\\
  c_{FAH} &=& -0.098(0.051)\,,
  \nonumber\\
  c_{FHH} &=& -\frac{427}{162} + \frac{158}{2835}\pi^2 +
  \frac{16}{9}\zeta(3)\,.
  \label{eq::cv3}
\end{eqnarray}
All uncertainties originating from the individual master integrals are added
quadratically.  In order to obtain a conservative error estimate we interpret
the uncertainty of the numerical integration as one standard deviation from a
Gaussian distribution and multiply it by a factor of five which is accounted
for in Eq.~(\ref{eq::cv3})~\footnote{Note that in
  Ref.~\cite{Marquard:2009bj}, where the fermionic contributions are given,
  only a factor two has been chosen which explains the slight increase of the
  uncertainty of $c_{FAH}$ in Eq.~(\ref{eq::cv3}).}.  The coefficients of
$\Lmu$ could be obtained in analytic form since all renormalization constants
are known analytically.

In most applications it is sufficient to know the result for the matching
coefficient with numerically evaluated color factors. Setting $C_F=4/3$,
$C_A=3$, $T=1/2$ and $n_h=1$ before inserting the master integrals and
combining the numerical uncertainties we get
\begin{eqnarray}
  c_v &\approx& 1 - 2.667 \frac{\alpha_s^{(n_l)}}{\pi}
  + \left(\frac{\alpha_s^{(n_l)}}{\pi}\right)^2\left[
    - 44.551 + 0.407 n_l \right]
  \nonumber\\&&\mbox{} 
  + \left(\frac{\alpha_s^{(n_l)}}{\pi}\right)^3\left[
    -2091(2)  +120.66(0.01)\, n_l 
  \right.\nonumber\\&&\left.\mbox{} 
    -0.823\, n_l^2
    \right]
  + \mbox{singlet terms}
  \,,
  \label{eq::cv3num}
\end{eqnarray} 
where $\mu=m_Q$ has been chosen.
Note that the $n_l$-independent three-loop term contains the contribution
with a closed massive quark loop which amounts to
$c_v^{(3)}|^{n_l=0}_{n_h}\approx -0.93(8)$~\cite{Marquard:2009bj}.
One observes that for $n_l=3,4$ and $5$ 
all coefficients in Eq.~(\ref{eq::cv3num})
have the same sign and that they grow
quite rapidly when going from NLO to NNNLO. At NNLO and NNNLO the fermionic
corrections screen the non-fermionic ones, but even for $n_l=5$ only a
reduction of at most 30\% is obtained.
A first glance at Eq.~(\ref{eq::cv3num}) would suggest that for
the quantity $c_v$ perturbation theory breaks down even though the momentum
scale 
involved in the problem, $m_Q$, is quite large. However,
as already mentioned above, $c_v$ itself does not represent a physical
quantity. It has to be combined with contributions originating from soft,
potential and ultrasoft momentum regions which can compensate the large
coefficients in Eq.~(\ref{eq::cv3num}). Further discussions on this topic
can be found in Ref.~\cite{bottom_NNNLO}.
It might very well be that the $\overline{\rm MS}$ scheme
adopted in our calculation is not well suited for separating the divergences
occuring in the different regimes. In fact, also the ultrasoft
contribution studied in Ref.~\cite{Beneke:2007pj,Beneke:2008cr} shows large
numerical effects. 

We have performed several checks on the correctness of our result which we want
to mention in the following. In our calculation we allowed for a general gauge
parameter $\xi$ which manifests as a polynomial dependence of the individual
diagrams. After summing the three-loop results for $Z_2$ and $\Gamma_v$ (taking
into account the corresponding quark mass counterterm contribution)
we concentrated on the coefficient of the linear $\xi$-dependence and 
have verified that it vanishes.
As a further check we recomputed the $n_l$ contribution~\cite{Marquard:2006qi}
using our automated setup. In this context we want to mention that in
Ref.~\cite{Marquard:2006qi} all occuring master integrals have been computed
either analytically or using a numerical method different from the
one used in the present paper. As already mentioned above, with our
calculation we could also reproduce the renormalization
constant in Eq.~(\ref{eq::Zv}) with high accuracy which checks all
but the highest $\epsilon$ coefficients of the master integrals.
We note in passing that we have a similar accuracy for the cancellation of the 
spurious poles up to seventh order occuring due to our
reduction procedure.

\begin{figure}[t]
  \includegraphics[width=.7\columnwidth]{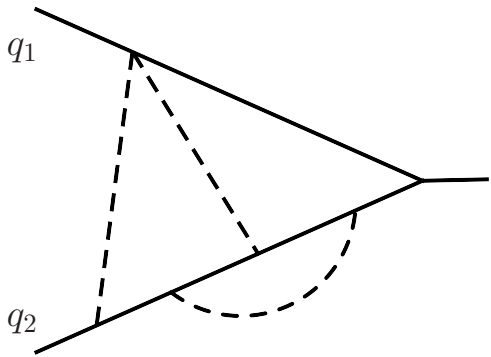}
  \\[-13em]
  \hspace*{-20em}
  \begin{minipage}{1em}
    \Large
    $q_1$

    \vspace*{6em}

    $q_2$
  \end{minipage}
  \\[2em]
  \caption{\label{fig::MI} Typical master integral appearing in our
    calculation. The solid lines and dashed lines represent massive and
    massless lines, respectively. For the external momenta we have the
    conditions $q_1^2=q_2^2=m_Q^2$ and $(q_1+q_2)^2=4m_Q^2$.
    }
\end{figure}

At this point it is instructive to show a result for a typical master integral
contributing to $c_v$. For the Feynman diagram in Fig.~\ref{fig::MI}, which we
need up to order $\epsilon$,
we obtain with the help of {\tt
  FIESTA}~\cite{Smirnov:2008py,Smirnov:2009pb,Smirnov:2013eza} 
\begin{eqnarray}
  M&=&\frac{e^{3\epsilon\gamma_E}}{m_Q^4}\left(\frac{\mu^2}{m_Q^2}\right)^{3\epsilon}
  \Bigg(
    +\frac{0.411236(3)}{\epsilon^2}
    +\frac{3.4860(1)}{\epsilon}
    \nonumber\\&&\mbox{}
    +34.520(2) 
    +339.68(4) \epsilon
    +{\cal O}(\epsilon^2)
  \Bigg)\,.
\end{eqnarray}

A very powerful check on the correctness of our result is provided by the
change of basis for the master integrals. We employ the integral tables
generated during the reduction procedure in order to re-express the master
integrals, which are not known analytically, 
through different, in general more complicated ones. This
transformation is done analytically for general space-time dimension $D$. In a
next step the new master integrals are again evaluated with {\tt FIESTA} and
inserted in the new expression for $c_v$. In Tab.~\ref{tab::basis} we compare
the results for the purely gluonic coefficients and the complete result
for ${c}_v^{(3)}$ obtained in the two bases. We observe an excellent
agreement within the uncertainties. 
In the case of the ``alternative basis''
one has to keep in mind that the integrals to be evaluated numerically
are significantly more complicated which explains the larger uncertainties for the
coefficients in Tab.~\ref{tab::basis}.

\begin{table}[t]
  \begin{center}
  \begin{tabular}{c|c|c}
    & default basis (cf. Eq.~(\ref{eq::cv3}))& alternative basis\\
    \hline
${c}_{FFF}$ & $36.55(0.11)$ & $36.61(2.93)$ \\
${c}_{FFA}$ & $-188.10(0.17)$ & $-188.04(2.91)$ \\
${c}_{FAA}$ & $-97.81(0.08)$ & $-97.76(2.05)$ \\
${c}_v^{(3)}$ ($n_l=4$) & $-1621.7(0.4)$ & $-1621(23)$ \\
${c}_v^{(3)}$ ($n_l=5$) & $-1508.4(0.4)$ & $-1507(23)$ \\
  \end{tabular}
  \caption{\label{tab::basis}Comparison of the purely gluonic coefficients of
    Eq.~(\ref{eq::cv3}) and $c_v^{(3)}$ with $n_l=4$ and $n_l=5$
    for two different choices
    of the master integral basis. For convenience $\mu=m_Q$ has been adopted.
    The given uncertainties are obtained by combining the numerical
    uncertainties of each master integral contribution in quadrature.
    In contrast to Eqs.~(\ref{eq::cv3}) and~(\ref{eq::cv3num}) no factor 
    five has been introduced for this comparison.}
  \end{center}
\end{table}

As a further check on the numerical evaluation of the master integrals we have
used a different momentum assignment in the input for {\tt FIESTA}. As a
consequence different expressions are generated in intermediate steps leading
to different numerical integrations. The final results are in complete
agreement with Eq.~(\ref{eq::cv3num}).

\begin{figure}[t]
  \includegraphics[width=\columnwidth]{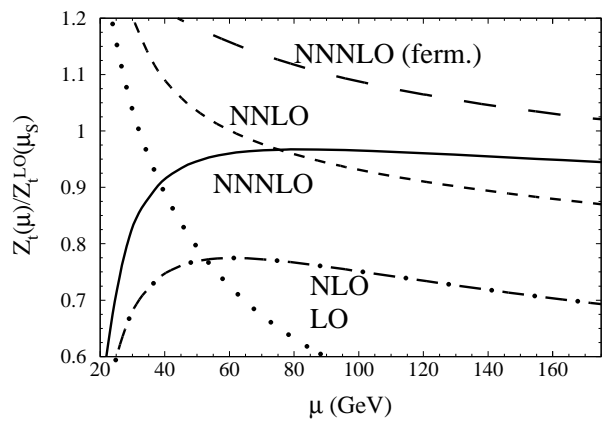}
  \caption{\label{fig::Zmu}Residue for the top quark system
    normalized to $Z_t^{\rm LO}(\mu_S)$
    as a function of the renormalization scale $\mu$.    
    Dotted, dash-dotted, short-dashed and solid lines correspond to LO, NLO,
    NNLO and NNNLO prediction. In the long-dashed curve only the
    fermion contributions to $c_v^{(3)}$ are taken into account 
    (NNNLO (ferm.)).
    }
\end{figure}

We are now in the position to have a first look to the phenomenological
consequences of our result for $c_v$. We 
consider the residue of the two-point function of the vector currents
\begin{eqnarray}
  \left(-q^2g_{\mu\nu}+q_\mu q_\nu\right)\,\Pi(q^2)
  &=&
  i\int {\rm d}x\,e^{iqx}\langle 0|Tj_\mu(x) j^{\dagger}_\nu(0)|0
  \rangle
  \,,
  \nonumber\\
  \label{eq::pivadef}
\end{eqnarray}
which is obtained by considering $\Pi(q^2)$ close
to the $Q\bar{Q}$ threshold. In this limit $\Pi(q^2)$ is dominated by
pole contributions originating from bound-state effects
\begin{eqnarray}
  \Pi(q^2) &\stackrel{E\to E_n}{=}& \frac{N_c}{2m_Q^2}\frac{Z_n}{E_n-(E+i0)}
  +\ldots
  \,,
\end{eqnarray}
where the ellipsis denotes contributions from the con\-ti\-nuum. 
$Z_n$ and $E_n$
are the residue and energy of the $n^{\rm th}$ resonance which determine the
height and position of the threshold cross section, respectively.  In the
following we consider the residue of the $1S$ (pseudo) bound state of
top quarks and extend the considerations of
Ref.~\cite{Beneke:2007uf}\footnote{Note that there is a misprint in Eq.~(5) of
  Ref.~\cite{Beneke:2007uf}: the term $E(1-d_v/3)/m$ should read
  $E(c_v-d_v/3)/m$.} by including the non-fermionic contribution of
$c_v^{(3)}$ and the ${\cal O}(\epsilon)$ term of the $1/m_Q$
potential~\cite{bottom_NNNLO}. Choosing the potential subtracted
scheme~\cite{Beneke:1998rk} with $\mu_f=20$~GeV
to define the top quark mass we obtain $m_t^{\rm PS}=171.4$~GeV which leads to
\begin{eqnarray}
  Z_t &=& \frac{(C_F m_t^{\rm PS} \alpha_s)^3}{8\pi}
  \left[ 1+\left(-2.131+3.661L\right)\alpha_s
    + \left(8.38
    \right.\right.\nonumber\\&&\left.\left.\mbox{}
      +1.27x_f-7.26\ln\alpha_s-13.40L+8.93L^2\right)\alpha_s^2
    \right.\nonumber\\&&\left.\mbox{}
    +\left(5.46 + \left(-2.23 + 0.78L_f\right)x_f
      + 2.21_{a_3} 
      \right.\right.\nonumber\\&&\left.\left.\mbox{}
      + 21.48_{b_2{\epsilon}} 
      + 37.53_{c_{f}}
      - 134.8(0.1)_{c_g}
      \right.\right.\nonumber\\&&\left.\left.\mbox{}
      +\left(-9.79
        -44.27L\right) \ln\alpha_s
      -16.35\ln^2\alpha_s
    \right.\right.\nonumber\\&&\left.\left.\mbox{}
      + \left( 53.17 + 4.66x_f \right)L
      -48.18L^2+18.17L^3\right)
    \alpha_s^3
    \right.\nonumber\\&&\left.\mbox{}
    + {\cal O}(\alpha_s^4)
  \right]
  \nonumber\\
  &=& \frac{(C_F m_t^{\rm PS} \alpha_s)^3}{8\pi}
  \left[1-2.13\alpha_s
    + 23.66\alpha_s^2
  \right.
  \nonumber \\&&\mbox{}
  \left.
    - 113.0(0.1)\alpha_s^3
    + {\cal O}(\alpha_s^4)
  \right]
  \,,
\end{eqnarray}
where $x_f=\mu_f/(m_t^{\rm PS}\alpha_s)$, $L = \ln\,(\mu/(m_t^{\rm PS}
C_F\alpha_s))$, and $L_f=\ln(\mu^2/\mu_f^2)$.
We have used $\alpha_s(M_Z)=0.1184$ to compute
$\alpha_s=\alpha_s(\mu_S)\approx 0.141$ where the soft scale
$\mu_S=m_Q C_F\alpha_s(\mu_S)\approx 32.16$~GeV has been
adopted after the second equality sign. In order
to get an impression about the importance of the individual contributions we 
mark the $\mu$-independent coefficients 
from the three-loop static potential ($a_3$), from
the two-loop ${\cal O}(\epsilon)$ term of the $1/(m_Qr^2)$ potential
($b_2\epsilon$), and from
the three-loop fermion ($c_{f}$) and purely gluonic ($c_{g}$)
contribution to $c_v^{(3)}$ separately.
For this choice of $\mu$ one observes  quite big NNNLO contributions which are
dominated by $c_{g}$. Thus, it is instructive to investigate the
$\mu$-dependence of $Z_t$ which is shown in Fig.~\ref{fig::Zmu}.
Around the soft scale  no convergence is observed.
Allowing, however, for higher scales one finds a quite flat behavior of the 
NNNLO curve. Furthermore, the NNNLO corrections become quite small.
E.g., considering the top quark system for $\mu\approx 80$~GeV, the NLO terms
amount to about $+15$\% and the NNLO to roughly $+20$\%. The third-order
contribution is practically zero.
Similar observations also hold for the bottom quark case, 
see Ref.~\cite{bottom_NNNLO}.


\medskip

\section{\label{sec::concl}Conclusions}

The third-order contribution to the matching coefficient of the vector current
between QCD and NRQCD has been computed. An automated setup has been
developed where even the occuring master integrals are identified
automatically, processed with the help of the computer program {\tt FIESTA},
and prepared for the insertion into the analytic reduction of $c_v^{(3)}$.

In the $\overline{\rm MS}$ scheme the numerical impact of $c_v^{(3)}$ is quite
big as can be seen from Eq.~(\ref{eq::cv3num}) which constitutes the main
result of this 
paper. In a dedicated analysis one has to investigate the consequences for the
bottom-quark mass extracted from $\Upsilon$ sum rules and the top-quark
threshold production cross section at a future linear collider.

An analysis of the residue of the $1S$ state indicates that at energy scales
around two to three times the soft scale good convergence of the perturbative
series is observed.




\begin{acknowledgments}
We would like to thank Martin Beneke and 
Alexander Penin for many useful discussions and
communications and Alexander Smirnov for continuous 
support on {\tt FIESTA}.
This work is supported by DFG through SFB/TR~9
``Computational Particle Physics''.
The Feynman diagrams were drawn with the help of
{\tt Axodraw}~\cite{Vermaseren:1994je} and {\tt
  JaxoDraw}~\cite{Binosi:2003yf}.
\end{acknowledgments}




\end{document}